\documentclass{aa}  

\usepackage{graphicx,natbib,txfonts}
\begin{document}

   \title{AGN mass estimates in large spectroscopic surveys: the effect
     of host galaxy light.}

%   \subtitle{I. Overviewing the $\kappa$-mechanism}

   \author{Ludovica Varisco\inst{1}, Tullia Sbarrato\inst{1}, Giorgio Calderone\inst{2}, 
                \and Massimo Dotti\inst{1,3}}

   \institute{Universit\`a degli Studi di Milano--Bicocca, Dipartimento di
                Fisica ``G. Occhialini'', Piazza della Scienza 3, 20126, Milano, Italy
                \email{l.varisco4@campus.unimib.it}
                \and  INAF - Osservatorio Astronomico di Trieste, via Tiepolo 11, I-34131, Trieste, Italy
                \and INFN, Sezione di Milano--Bicocca, Piazza della Scienza 3, 20126, Milano, Italy
             }

   \date{}

% \abstract{}{}{}{}{} 
% 5 {} token are mandatory
 
  \abstract
% context heading (optional)   {}
  % aims heading (mandatory)   {}
  % methods heading (mandatory)   {}
  % results heading (mandatory)   {}
  % conclusions heading (optional), leave it empty if necessary    {}
  {Virial--based methods for estimating active supermassive black hole masses  
  are now commonly used on extremely large spectroscopic quasar catalogs. 
  Most spectral analyses, though, do not pay enough attention 
  to the detailed continuum decomposition. 
  To understand how this affects virial mass estimates, we test the influence 
  of host galaxy light on them, along with a Balmer continuum component. 
  A detailed fit with the new spectroscopic analysis software {\sc QSFit} demonstrates 
  that the presence or absence of continuum components 
  does not significantly affect the virial--based results for our sample. 
  Taking a host galaxy component into consideration or not, instead, affects 
  the emission line fitting in a more pronounced way at lower redshifts, 
  where in fact we observe dimmer quasars and more visible host galaxies. 
  }
   

   \titlerunning{Host galaxy influence on AGN mass estimates}
   \authorrunning{L. Varisco et al.}

   \maketitle
%
%-------------------------------------------------------------------

\section{Introduction}

Massive black holes (MBHs), commonly observed in the centers of massive
galaxies, can be fully characterized by only two parameters: their masses and
their spins \citep[see e.g.,][and references therein]{sesana14}. Since MBH
spins affect the dynamics of matter only within their innermost few
gravitational radii, spins are usually constrained through observationally
challenging X-ray spectroscopy complemented with relativistic modeling of
broad iron lines \citep[see][for recent reviews on the topic\footnote{
    But see \cite{davies11, trakhtenbrot14, capellupo16,
    campitiello18} for alternative techniques.}]{brenneman13,
  reynolds14}.  On the other hand, MBH masses
can dominate the dynamics of gas and stars up to parsec scales \citep[depending on the
MBH masses and the host properties; see e.g.,][]{dotti12} and are
therefore easier to constrain. Direct measurements are possible in a
limited number of cases, through orbital fitting of single stars as for the
case of our Galaxy \citep[see e.g.,][]{genzel10,meyer12,gillessen09a, gillessen09b}, 
through fitting the dynamics of megamaser
emitting disks \citep[][]{braatz97,greene10,vandebosch16}, or by modeling the dynamics of whole
stellar and/or gaseous populations in the immediate vicinities of the MBHs 
\citep[e.g.,][for a review]{gebhardt03,vandermarel98,ferrarese05}.

Alternative options for estimating MBH masses are available for the
subclass of Type I active MBHs (AGN). 
A widely used method to perform MBH mass estimates for this class of sources is based on the virial theorem \citep[see][and references therein]{peterson14}. The main assumption in this approach is that the gas in the sub-parsec region emitting the broad lines (BLR) is virialized 
and its dynamic is dominated by the central MBH. Therefore the virial MBH mass is determined by:
\begin{equation}
M_{\rm vir}= \frac{V^2 R_{\rm BLR}}{G} = \alpha \, \frac{\mathrm{FWHM}^2 \, L_{\rm cont}^{\beta}}{G}
,\end{equation}
where $V$ is the virial velocity and $R$ the size of the BLR. 
Assuming that broad emission lines (BELs) are Doppler broadened due to 
the virial motion of the BLR, the full width at half maximum (FWHM) of the BELs can be used as a proxy of the
 emitting gas velocities.
Moreover, through an analysis of the time-delayed response of BELs to variations 
of the continuum, reverberation mapping campaigns allow us to estimate the size of the BLR.  
Reverberation mapping studies have also revealed the so--called size--luminosity relation between the 
monochromatic luminosity of the AGN continuum and the BLR size \citep[][]{kaspi00,bentz06}. 
Continuum luminosity can therefore be used as a proxy of the BLR size.
Hence, MBH mass can be estimated by means of a fudge factor $\alpha$ that includes uncertainties
 on BLR shape and dynamics \citep[see e.g.,][]{decarli11}. $\alpha$ factor can however be constrained self-consistently via reverberation mapping studies in cases where 
 extremely high-quality and high-coverage data are available \citep[][]{brewer11,pancoast12}. 

 \cite{shen11} presented a systematic study of the SDSS DR7 Quasar Catalogue \citep[][]{schneider10}, in which every spectrum is automatically fitted with a power--law continuum, 
along with broad and narrow emission lines and an iron template. MBH masses were constrained from the FWHM of different BELs (depending on the
AGN redshift), and the monochromatic luminosity in the proximity of
the BEL used. In this paper we re--examine all the AGN in the
SDSS--DR10 catalog by performing an independent fit of each AGN
spectrum with the publicly available {\sc QSFit} code
\citep{calderone17}\footnote{The code can be downloaded from the
  http://qsfit.inaf.it webpage.}. In addition to the spectral features
related to the accreting MBH, the code includes an elliptical galactic
template \citep{polletta07}. This feature is important for two reasons: ($i$) it could
contribute to the monochromatic luminosity, resulting in the
overestimation of single epoch MBH masses, and ($ii$) depending on its
intensity and spectral shape, it could affect the BEL shape and,
hence, the FWHM.  The contamination from the host galaxy can be
drastically reduced in multiple ways, for example, by taking high-angular-resolution nuclear spectra, or by considering only the varying
component of the emission from multi-epoch spectral studies of the
same AGN (as done in reverberation mapping). We stress however that
none of these approaches can be applied to very large AGN samples
characterized by broad distributions of AGN luminosities and
redshifts, such as the SDSS catalog. Therefore, in order to maximize
the payoff of such large surveys, the fitting of the host component is
mandatory.
Some methods have been proposed in the literature to estimate the contamination due to
 the host galaxy in continuum luminosity measurements. \citet{shen11}  compute a statistical analysis on low--redshift sources (in which we expect
a higher host--galaxy contamination) and provide an empirical fitting formula. In their work the resulting host galaxy contamination is on average $\sim$ 15\% of the 
the continuum luminosity at 5100 \AA\ on low--redshift sources ($z\lesssim 0.5$).
Other authors proposed different approaches such as the use of the equivalent width of the 
line Ca $\rm II$ K line at $\rm 3934 \AA$ as a proxy of the degree of host contamination \citep{greene05}. 

Different single epoch relations have been proposed to remove the
 host galaxy contamination \citep[e.g., Eq. 6 and 7 of ][]{greene05,bongiorno14},
 based on empirical relations observed on large samples of AGNs. Hence these relations
 are only able to estimate host galaxy contributions on average, not on individual
 sources.
In this work, we analyzed each source individually to isolate the host galaxy contribution and
 avoid any possible bias introduced by the choice of the sample for the empirical 
 relations.

The paper is organized as follows: in section 2 we discuss how we select the
AGN sample used and review the main features of the spectral
fitting implemented in {\sc QSFit}. In section 3 we discuss our results and
how they depend on the redshift and on the AGN properties. Finally, in section 4
we draw our conclusions and discuss possible future improvements.

\section{Sample selection and data analysis}
\label{sec:analysis}
Our prime sample was composed of the 10578 sources in the {\sc QSFit}
catalog \citep[version 1.2][]{calderone17} with $z < 0.8$ for which a
measurement of the AGN continuum luminosity at 5100 $\rm \AA$ and of
the $\rm FWHM_{H\beta}$ was available in Shen's catalog
\citep{shen11}. We analyzed the SDSS--DR10 optical spectra using the
      {\sc QSFit} software \citep[version 1.2][]{calderone17}.  Each
      spectrum is de--reddened following the prescription by
      \citet{cardelli89} and \citet{odonnell94}, then transformed to
      rest frame and finally analyzed using the {\sc QSFit} recipe
      which simultaneously fits all the spectral components visible in
      the observed wavelength range. {\sc QSFit}  uses a power law to model the
      AGN broad band, a component to account for the Balmer continuum
      \citep[following the recipe by][]{grandi82,dietrich02}, the
      \citet{polletta07} elliptical galaxy template to account for the
      host galaxy emission, the iron templates by
      \citet{vestergaard01} and \citet{veron04} at UV and optical
      wavelengths respectively, and several Gaussian profiles to
      account for the most commonly observed emission lines.  We
      collected all the output values and extracted the relevant
      quantities for the purpose of this work, namely the $\lambda
      L_\lambda$ continuum luminosity at 5100\AA{} ($L_{5100}$), the
      Balmer continuum contribution, and the FWHM of the H$\beta$
      emission line. 

The inclusion of the
Balmer continuum component in the fit procedure affects the slope of the
continuum and could, as a consequence, modify the continuum at longer
wavelengths. By running the analysis with and without considering the
Balmer continuum component (by setting the {\tt !QSFIT\_OPT.balmer}
      option to 0) we quantify its impact on the MBH mass estimates.

We selected the sources for which the fit quality flag of the continuum at 
5100 $\rm \AA$  and of the H$\beta$ broad emission line components were good 
(namely {\tt CONT5100\_QUALITY} and {\tt br\_Hb\_QUALITY} 
simultaneously equal to 0) for both standard and without Balmer component analysis.
The final sample consisted of 9107 sources. 

In the following, results of the standard analysis will be flagged as $L_{5100}^{\rm B}$
and FWHM$_{\rm H\beta}^{\rm B}$, while the quantities obtained by 
disabling the Balmer continuum component will be flagged as 
$L_{5100}^{\rm nB}$ and FWHM$_{\rm H\beta}^{\rm nB}$.

   \begin{figure}
   \centering
   \includegraphics[width=\hsize]{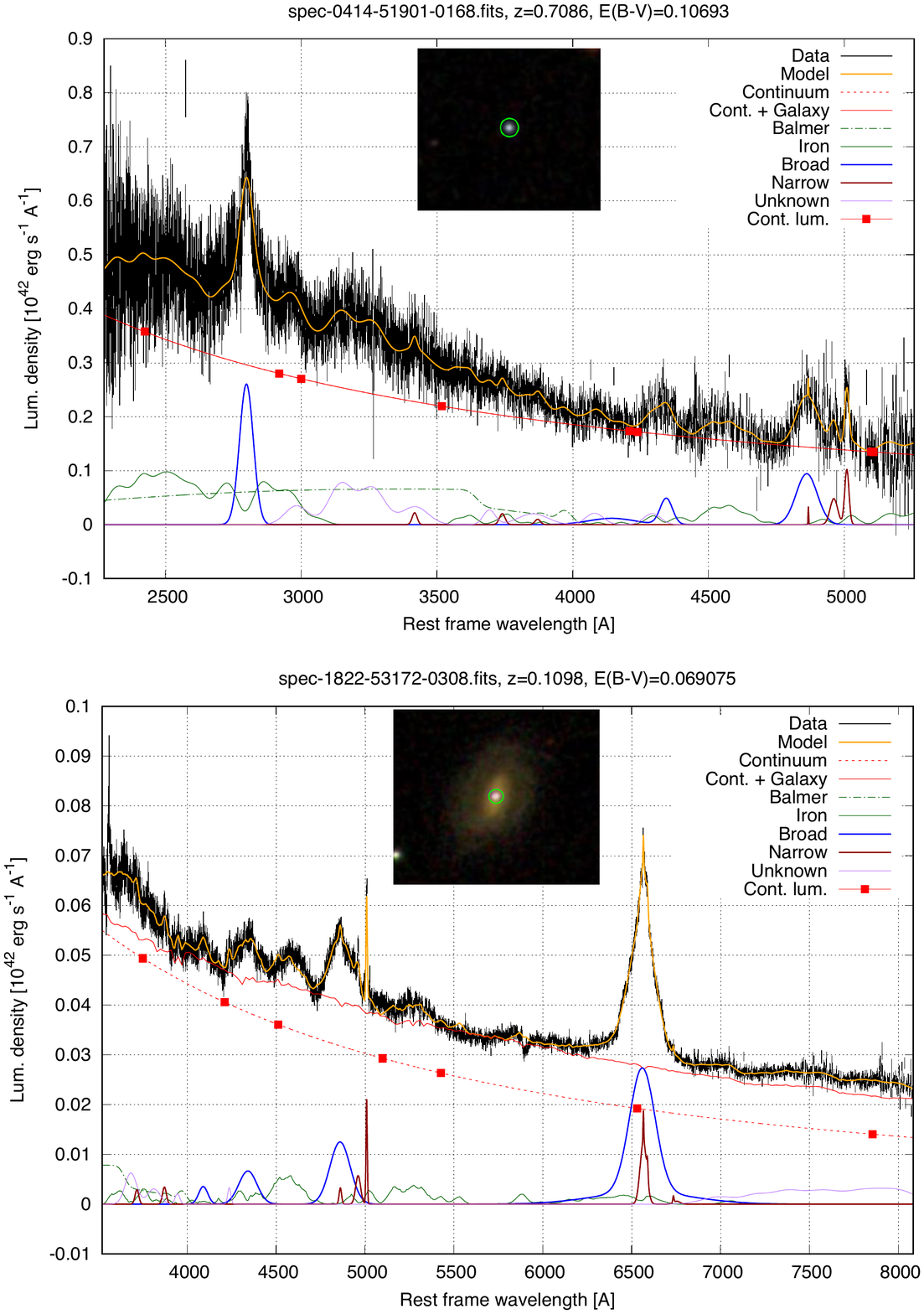}
      \caption{SDSS spectra and images of SDSS J032545.70-000820.6 ($z
        \simeq 0.7$; \textit{upper panel}) and SDSS
        J155053.16+052112.1 ($z \simeq 0.1$; \textit{lower panel}).
        In each spectrum, SDSS data with respective errors (black
        lines) are compared with the {\sc QSFit} model (orange solid
        line). The red dashed line is the continuum component, while
        the red solid line includes both AGN and host galaxy
        continuum. The dot-dashed green line is the Balmer continuum
        component; the green solid line is the iron template.  Broad
        emission lines are shown with solid blue lines, while narrow
        emission lines with solid brown lines. The solid purple line
        is the sum of all the "unknown" lines. The spatial region from which the
         SDSS spectra were obtained is shown in the 
         SDSS images with a green circle with a diameter of 3 arcsec.
         The two sources have
        significantly different contributions from the host galaxy:
        this is evident both in the SDSS images and in the spectra.
       In our fitting procedure we did not fix the FWHM and fluxes ratio
         of the [OIII] doublet. This affects individual features of the [OIII] 
        doublet, but does not affect the mass estimates.
          }
         \label{Fig1}
   \end{figure}

\section{Supermassive black hole mass estimates}
\label{sec:results}

%The main focus of this work is the influence of the host galaxy light
%on the supermassive black hole mass estimates. Therefore we aim at
%comparing virial supermassive black hole estimates obtained from large
%spectroscopic catalogs, to virial results calculated with a more
%detailed spectroscopic analysis.  On one side, we consider the results
%obtained by S11 in their broad spectroscopic study.  In fact, S11 is
%currently the most used spectroscopic quasar catalog that provides
%black hole mass estimates of large AGN samples.  Though, they did not
%include a component to model the host galaxy emission, and might
%therefore have overestimated the AGN continuum luminosity, affecting
%the supermassive black hole estimate.  On the other hand, {\sc QSFit}
%spectroscopic analysis includes a host galaxy component, therefore in
%the following we will compare our results with S11, in order to derive
%the host contribution to black hole mass estimates.

Figure \ref{Fig1} shows the results of our full spectral fitting
procedure (including the Balmer continuum component; see the
discussion above) performed over two AGN-host galaxy systems,
exemplifying the different possible degrees of host galaxy
contribution to SDSS spectra.  The upper panel shows the SDSS image
and spectrum of SDSS J032545.70-000820.6, a source with a host galaxy
significantly fainter than the AGN, while the lower panel shows SDSS
J155053.16+052112.1, an AGN hosted in a brighter galaxy.  Other than
being clearly different in the SDSS images (host galaxy emission almost unresolved in the upper panel), the two nuclei
significantly differ in their spectra. The spectrum of SDSS
J032545.70-000820.6 is well fitted without any need for a relevant
host galaxy contribution. On the other hand, the analysis performed
on the spectrum of SDSS J155053.16+052112.1 reveals a dominant
contribution of the host galaxy light in the continuum emission. In
this case an analysis performed without considering the host galaxy
contribution would lead to an overestimate of at least a factor 1.3 in
the 5100 \AA\ continuum 
\footnote{It corresponds to a host galaxy contribution of $\sim 25 \%$, significantly larger than the $\sim 15 \%$ estimated by \citet{shen11}.}.
A poor estimate of the continuum could also
affect the emission line profile reconstruction: S11 fits BELs with a
combination of up to four Gaussian profiles in order to account for
asymmetries, while in 82\% of cases we require only
  one component to achieve a good fit
 \footnote{When multiple Gaussians are needed to achieve a good fit, the 
 FWHM is derived from the total line profile.}.
 
% \textbf{Figure \ref{Lcomparison} shows the comparison between the 
% $L_{\rm 5100,host}-L_{\rm 5100,agn}$ ratio proposed 
% by \citet{shen11} with an empirical formula computed with S11 data 
%  and the one computed with {\sc QSFit} data. }

%
%   \begin{figure}
%   \centering
%   \includegraphics[width=\hsize]{plots/L5100comparison.png}
%      \caption{\textbf{comparison between the $L_{\rm 5100,host}-L_{\rm 5100,agn}$ ratio proposed 
%      by \citet{shen11} with an empirical formula computed with S11 data 
%      and the one computed with {\sc QSFit} data. 
%      The black dashed line is the equality line.}
%          }
%         \label{Lcomparison}
%   \end{figure}

After running the two variants of spectral fitting discussed in
Sect.~\ref{sec:analysis} over our entire sample, we proceed to estimating the MBH masses ($M$) using the two following calibrations:
\begin{equation}\label{eq:mass1}
  M = 0.78 \times10^8 \left(\frac{L_{5100}}{10^{46}{\rm erg/s}}\right)^{0.61} \left(\frac{\rm FWHM_{H\beta}}{1000{\rm \,km/s}}\right)^2 {\rm M_\odot},
\end{equation}
originally proposed by \citet{mclure04}, and
\begin{equation}\label{eq:mass2}
M = 0.813 \times10^8 \left(\frac{L_{5100}}{10^{46}{\rm erg/s}}\right)^{0.5} \left(\frac{\rm FWHM_{H\beta}}{1000{\rm \,km/s}}\right)^2 {\rm M_\odot},  
\end{equation}
proposed by \citet{vestergaard06}. Neither calibration includes the host galaxy contribution.
 These relations have been chosen to compare our
findings with the estimates presented in S11 in order to quantify the
effect of the host galaxy light on the estimate of $M$. The masses
obtained from the full fitting procedure (i.e., the masses obtained
using $L_{5100}^{\rm B}$ and FWHM$_{\rm H\beta}^{\rm B}$ in
Eqs.~\ref{eq:mass1}~and~\ref{eq:mass2}) are dubbed $M_{\rm
  QSFit}^{\rm B}$, while with $M_{\rm QSFit}^{\rm nB}$ we indicate the
mass estimates from the fitting that does not include the Balmer
contribution.
%{\bf manca un riferimento a fig 2... Non sono sicura di ricordarmi 
%a cosa ci servisse: se fosse per sottolineare l'importanza della 
%galassia nella stima di FWHM, in questo caso non si vede.. 
%Che ne dite di toglierla?}
%
%   \begin{figure}
%   \centering
%   \includegraphics[width=\hsize]{plots/Hb_zoom.png}
%      \caption{Spectrum of SDSS J085404.83+193341.9, with a zoom--in around its% $H\beta$ emission line. 
%                   }
%         \label{Fig2}
%   \end{figure}

\begin{figure*}
   \centering
   \includegraphics[width=0.9\hsize]{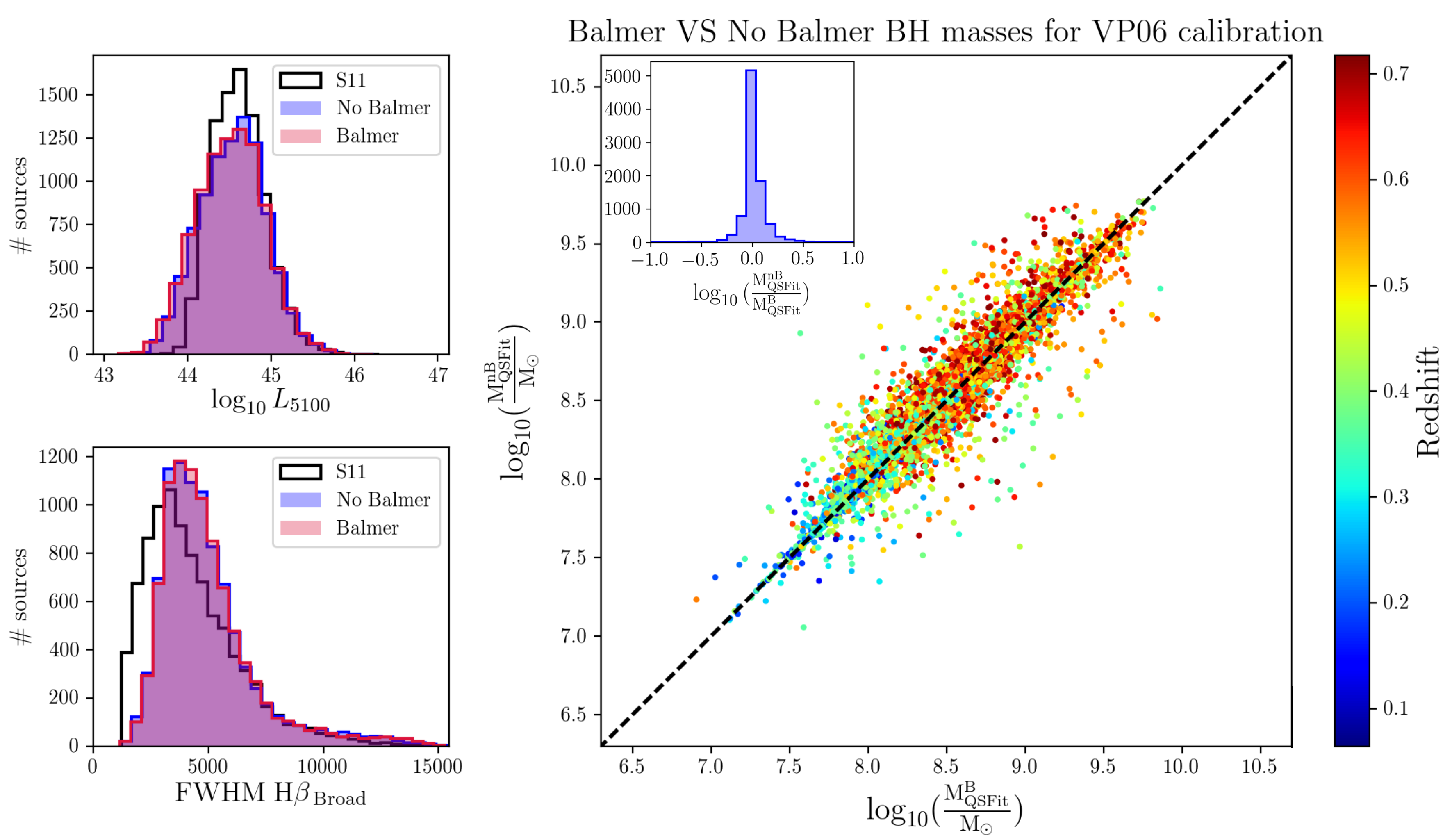}
      \caption{\textit{Left panels}: Distributions of $5100 \, \AA$ continuum luminosity
       ({\it upper}) and FWHM$_{\rm H\beta}$ ({\it lower}).
        Black histograms show results from S11.  Red and
        blue filled histograms are the {\sc QSFit} results obtained including
        and without the Balmer continuum component, respectively.
        \textit{Right panel}: Comparison between $M_{\rm QSFit}^{\rm B}$ and
        $M_{\rm QSFit}^{\rm nB}$.  The points are colored in redshift.
        The dashed black line is the equality line. 
        The blue histogram shows the distribution of the logarithmic ratio between 
        $M_{\rm QSFit}^{\rm nB}$ and $M_{\rm QSFit}^{\rm B}$.
        The three panels show that results obtained with or without the Balmer
        continuum component are substantially consistent. }
         \label{Fig3}
   \end{figure*}

Figure~\ref{Fig3} presents an overview of the results of our fitting
procedures.  In the upper left panel we show a comparison
between the distribution of 5100 \AA\ continuum luminosity measured by S11
($L_{5100}^{\rm S11}$) and the distributions of $L_{5100}^{\rm B}$ and
$L_{5100}^{\rm nB}$ obtained with the {\sc QSFit} analysis.  While the
parameters obtained with and without the Balmer component are consistent (see
also the other panels in Figure~\ref{Fig3} and the following discussion), the
$L_{5100}^{\rm S11}$ distribution clearly differs from the others, in that it is
completely missing the low-luminosity tail: the less luminous AGNs have more likely a
relevant host galaxy component, that contributes significantly to the
5100 \AA\ monochromatic luminosity.

The same comparison between the FWHM$_{\rm H\beta}$ distributions is shown in the
lower left panel. Again, the FWHM distributions obtained with or
without including the Balmer component in the {\sc QSFit} analysis are
comparable, while S11 estimate systematically narrower FWHM. Given that
the {\sc QSFit} results obtained with or without Balmer continuum
component are substantially consistent, the resulting distribution of
$M_{\rm QSFit}^{\rm B}$ and $M_{\rm QSFit}^{\rm nB}$ is compatible, 
as shown in the right panel of
Fig.~\ref{Fig3}. As a consequence, in the following we limit
our comparison to the masses presented in S11 and those obtained
through the full {\sc QSFit} analysis ($M_{\rm QSFit}^{\rm B}$).

   \begin{figure*}
   \centering
   \includegraphics[width=0.48\hsize]{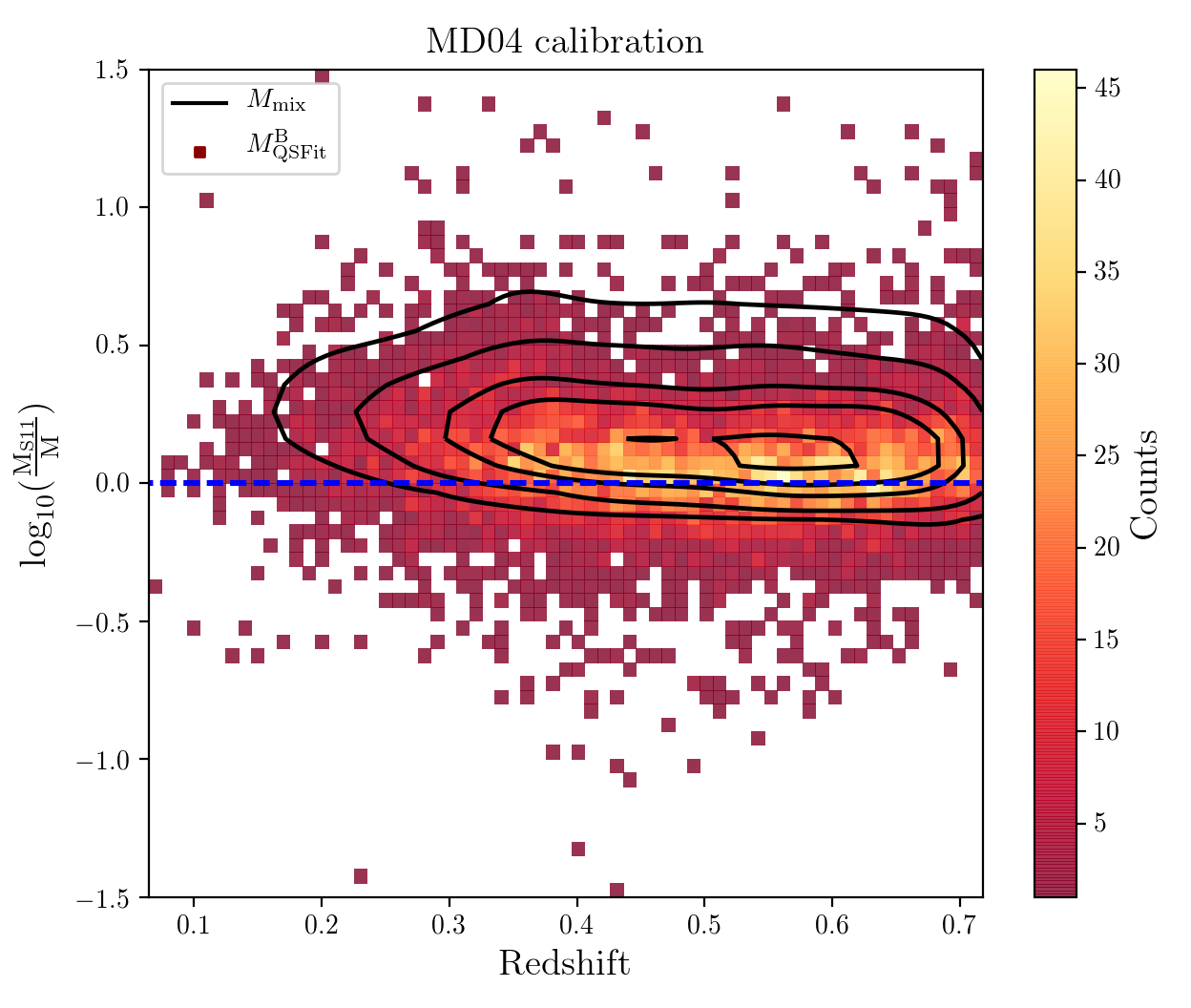}
   \includegraphics[width=0.48\hsize]{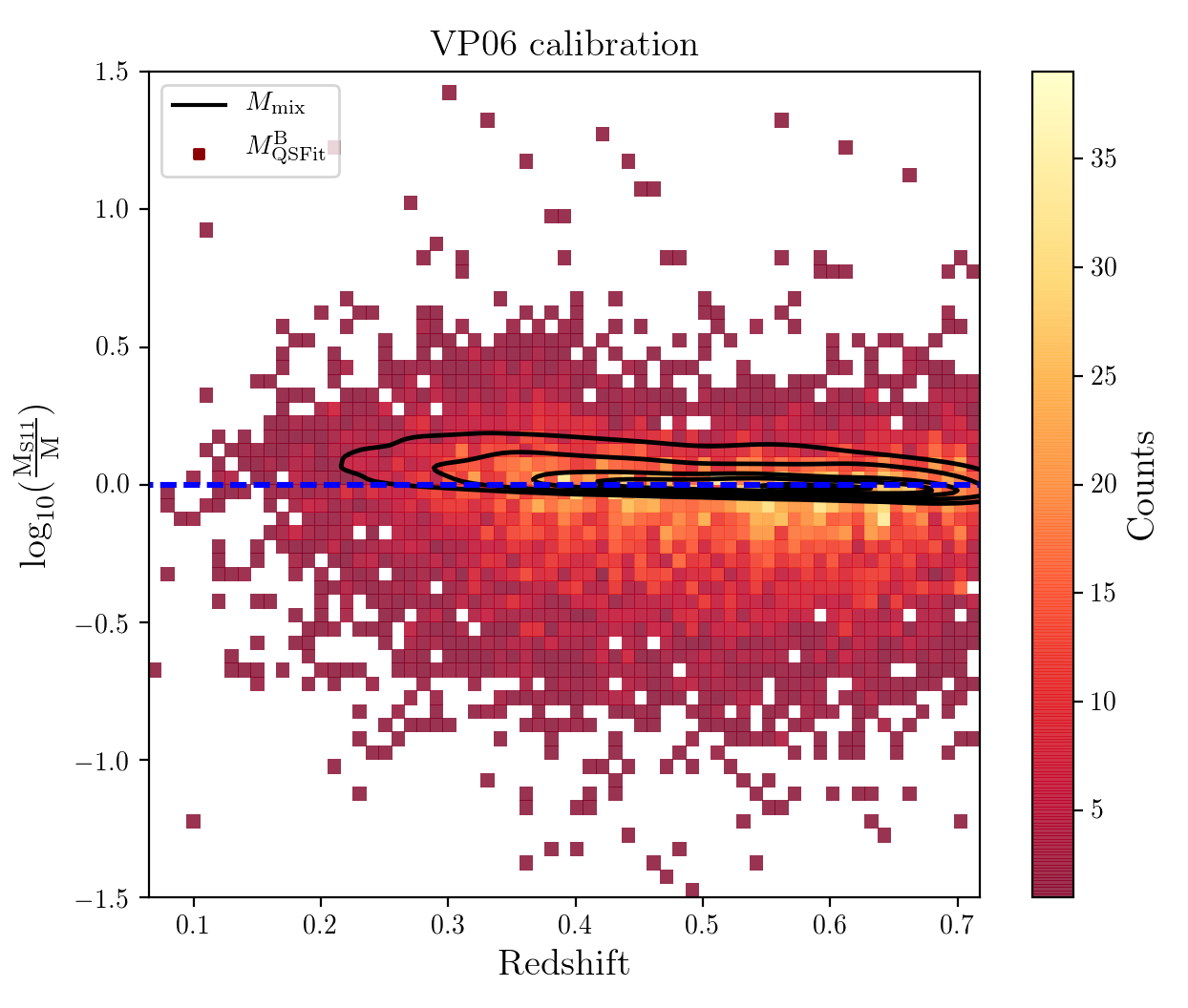}
      \caption{ Ratio of  virial masses of S11 and this work as a function
of redshift.  
      The color-coded two-dimensional histogram refers to ratios in which the MBH virial masses $
        M_{\rm QSFit}^{\rm B}$ have been obtained directly from the {\sc QSFit}
        results from Eq.~\ref{eq:mass1} (left panel) and
        Eq.~\ref{eq:mass2} (right panel).   
        The redshift is binned with a step of 0.01, while the S11-{\sc QSFit} mass ratio is binned
        with a bin step of 0.05.  
         The locus of the ratios obtained using $ M_{\rm mix}$ as denominator is mapped with the
        density contours, which are drawn at $10\%$, $25\%$, $50\%$,
        $68\%$ and $95\%$ of the distribution; with $M_{\rm mix}$ being the MBH mass calculated
        using S11 FWHM and our AGN continuum luminosity. For each calibration the blue
        dashed line shows the equivalence between our mass estimate and 
        $M_{\rm S11}$.}
         \label{final_fig}
   \end{figure*}

Figure~\ref{final_fig} shows the main result of our investigation. The two-dimensional 
color-coded histogram indicates the redshift distribution of the $M_{\rm
  S11}$ to $M_{\rm QSFit}^{\rm B}$ ratio. 
The different slope of the continuum luminosity at 5100 $\rm \AA$ and the different normalization in the two calibrations affect the ratio $M_{\rm S11}/M$ distributions shown with two-dimensional histograms in Fig. \ref{final_fig}. Indeed in Eq. \ref{eq:mass1} the $L_{5100}$ slope is 0.61 resulting in a $M_{\rm S11}/M$ ratio generally $\gtrsim 1$ for \citet{mclure04} calibration; while in Eq. \ref{eq:mass2} the $L_{5100}$ slope is 0.5 resulting in a $M_{\rm S11}/M$ distribution $\lesssim 1$ for \citet{vestergaard06} calibration.
  The two masses are comparable within a
factor of $\sim 3$ at all redshifts, but a large scatter is present, mostly due
to the different FWHM$_{\rm H\beta}$ estimates that enter quadratically in
Eqs.~\ref{eq:mass1}~and~\ref{eq:mass2}. This indicates how much the
single epoch virial estimates are affected by the inclusion (or not) of a host
galaxy component and by the decision on the number of Gaussian profiles used
to fit the BELs. As a second-order correction, on average, S11 estimates a
higher mass compared to {\sc QSFit} at low redshift, as highlighted by the
asymmetric distribution of the sources in Fig.~\ref{final_fig}. To investigate how a correct decomposition of
the observed continuum affects the virial mass estimates, we consider a
``mixed'' virial supermassive black hole mass estimate $M_{\rm mix}$,
calculated using S11 FWHM and our AGN continuum luminosity. The comparison
between $M_{\rm S11}$ and $M_{\rm mix}$ is also shown in
Fig.~\ref{final_fig} using density contours. Using the same value for the
FWHM$_{\rm H\beta}$ drastically reduces the scattering, while the redshift
dependence of the host galaxy contamination is emphasized: At lower
redshifts, where the host galaxy contribution is likely more relevant, the S11
result overestimates the virial mass calculated in this work. At high
redshifts, instead, the two mass estimates are comparable; quasars are likely
brighter and the host galaxy contribution is much more diluted in the
continuum emission.

%--------------------------------------------------------------------

%-------------------------------------- Two column figure (place early!)

%
   
%--------------------------------------------------- One column table
 
%
% 
%                                                One column figure
%----------------------------------------------------------------- 
%   \begin{figure}
%   \centering
%   %%%\includegraphics[width=3cm]{empty.eps}
%      \caption{Vibrational stability equation of state
%               $S_{\mathrm{vib}}(\lg e, \lg \rho)$.
%               $>0$ means vibrational stability.
%              }
%         \label{FigVibStab}
%   \end{figure}
%-----------------------------------------------------------------

\section{Conclusions}

We present the results of an automated spectral analysis of a
subsample of the SDSS DR10 quasar catalog. Our analysis, performed using
the public code {\sc QSFit}, fits the SDSS spectra including a host galaxy
template together with the AGN components (power-law continuum, broad and
narrow emission lines, and a Balmer). We used the
broad FWHM$_{\rm H\beta}$ and the monochromatic luminosity at 5100 $\rm \AA$ to
estimate the MBH masses responsible for the nuclear emission. 

We demonstrated that the inclusion of the Balmer
does not significantly affect the H$\beta$ mass estimates in our sample, while
the inclusion of a host galaxy component can significantly change both the BEL
shape and the 5100 $\rm \AA$ AGN monochromatic luminosity. In order to
quantify such an effect, we compared our estimates with those presented by
S11. We find a general agreement with S11, with the MBH mass estimates being
consistent within a factor of three. The main cause of the observed scattering
is due to the quadratic dependence of $M$ on the FWHM$_{\rm H\beta}$, the latter
being dependent on the continuum assumed and on the number of Gaussian
profiles allowed in the fit. Considering the intrinsic uncertainties associated
 with the single epoch estimators, which are of the order of $\sim 4$, our estimates are consistent with those of S11.  

Once the effect of the different FWHM of the broad H$\beta$ is removed
(using the same values measured by S11), the effect of including a
host galaxy template becomes evident. Depending on the calibrations
used, S11 either overestimates $M$ by about a factor of 2,
or, when the calibration from \citet{vestergaard06} is assumed
(Eq.~\ref{eq:mass2}), a significantly better agreement is
reached, with the masses becoming compatible within a factor of 1.5.
 In this last case, at low redshift, where the host galaxy
significantly contributes to the continuum luminosity, our mass
estimates are still slightly lower than those of S11. At higher
redshift ($z\gtrsim 0.6$), the estimates instead converge within 0.15 dex, the AGN overwhelming any contribution from the host galaxy.

%\begin{acknowledgements}
%
%\end{acknowledgements}


\end{document}